\documentclass[]{relaxed_system_lab}

\usepackage{url}
\usepackage{amsmath,amssymb}
\usepackage{pifont}
\usepackage{algorithm}
\usepackage{algpseudocode}
\usepackage{enumitem}
\usepackage{multirow}
\usepackage{wasysym}
\usepackage{xspace}

\newcommand{\system}{\textsc{Trident}\xspace}

\newtoggle{showmarks}
\toggletrue{showmarks}  %%%    <--- comment out this line :)

\iftoggle{showmarks}{
  \newcommand\zz[1]{\textcolor{red}{ZZ: #1}}
  \newcommand\TODO[1]{\textcolor{red}{TODO: #1}}
}{
  \newcommand\zz[1]{\unskip}
  \newcommand\TODO[1]{\unskip}
 }

\title{\system: Adaptive Scheduling for Heterogeneous Multimodal Data Pipelines}
\author{Ding Pan$^1$, Zhuangzhuang Zhou$^2$, Long Qian$^2$, Binhang Yuan$^1$}
\affiliation{$^1$HKUST, $^2$Huawei}

\abstract{
The rapid adoption of large language models and multimodal foundation models has created
unprecedented demand for high-quality training data, making multimodal data preparation
pipelines a critical AI infrastructure. These pipelines interleave CPU-heavy preprocessing
with AI accelerator-backed (e.g., GPU/NPU/TPU) inference operators and process massive intermediate artifacts.
Achieving high throughput is difficult for such a computation paradigm, where the multimodal workloads are highly
non-stationary: workload regime shifts and input-dependent inference behavior cause rapid
performance fluctuations, while transient memory spikes can trigger out-of-memory (OOM)
failures. Existing schedulers often rely on threshold-based autoscaling or assume
synchronous, homogeneous operators, leading to severe inefficiency.

To overcome these essential challenges, we introduce \textsc{\system}, an adaptive scheduling framework for heterogeneous
multimodal data pipelines on fixed-resource clusters, integrating three tightly coupled
layers in a closed loop:
(\underline{\textbf{i}}) an \emph{observation layer} that provides noise-resilient capacity estimates for asynchronous operators using Gaussian Process regression with two-stage anomaly filtering;
(\underline{\textbf{ii}}) an \emph{adaptation layer} that tracks workload regime shifts via online clustering and
uses memory-constrained Bayesian optimization to tune operator configurations
sample-efficiently, reducing OOM risk;
(\underline{\textbf{iii}}) a \emph{scheduling layer} formulates a mixed-integer linear program (MILP) that
jointly optimizes operator parallelism, placement, and configuration transitions under
heterogeneous compute and bandwidth constraints, trading off throughput gains against
cold-start overhead via rolling updates.
The three layers form a closed control loop:
capacity estimates and configuration recommendations flow into the MILP, whose decisions in turn trigger sample invalidation and model updates, ensuring system-wide consistency.
We evaluate \textsc{\system} on production-representative document and video curation
pipelines. \system improves end-to-end throughput by up to $2.01\times$ (PDF pipeline) and
$1.88\times$ (video pipeline) over a static baseline, substantially outperforming existing
schedulers, with low overhead suitable for online re-optimization.
}

\begin{document}
\maketitle
\section{Introduction}

The rapid advancement of large language models (LLMs) and multimodal foundation models (MFMs) has created an unprecedented demand for high-quality training data~\cite{compute-optimal,bai2025qwen3vltechnicalreport,run-out-of-data}. Multimodal data preparation pipelines that process petabytes of documents~\cite{redpajama,bai2025qwen3vltechnicalreport}, images~\cite{laion-5b}, and videos~\cite{opensora} have become a critical infrastructure for modern generative AI development. These pipelines contain various operators with heterogeneous resource utilization, such as video decoding, filtering operators using CPUs, alongside LLM-based assessment operators running on AI accelerators such as GPUs, TPUs, or NPUs, where a large volume of intermediate multimodal data flows through the pipelines.

We believe that stream processing architectures~\cite{spark-streaming,dataflow,flink,ray_data} should be able to offer compelling advantages for such workloads---unlike batch processing, which executes stage-by-stage and materializes all intermediate results between stages, stream processing divides data into dynamically sized batches that continuously flow between CPU preprocessing and GPU inference operators, and enables memory-efficient pipelining with concurrent utilization of heterogeneous resources. Emerging streaming data processing systems for AI and multimodal workloads (e.g., Ray Data~\cite{ray_data}) have demonstrated substantial throughput improvements over batch processing systems such as Spark~\cite{spark}. 

However, achieving high end-to-end throughput in production multimodal data pipelines remains challenging: \underline{First}, multimodal workloads are highly non-stationary: dataset skew and input dependent inference behavior~\cite{orca-contbatch,vllm} can cause large and rapid fluctuations in both performance and resource utilization during pipeline execution, including transient spikes in accelerator memory footprint that may trigger out-of-memory (OOM) failures. This requires the scheduler to model pipeline performance and make resource allocation decisions based on real-time performance metrics, while also accounting for memory safety. \underline{Second}, accurately characterizing the performance of multimodal data pipelines composed of complex asynchronous operators is essentially difficult. Operators often encapsulate LLM inference and batched vision models that execute asynchronously with complex internal scheduling logic, a vast configuration space, and shifting performance. Configuration changes that improve throughput (e.g., larger batches or longer sequences) can also push memory usage over the device limit and cause OOM-induced restarts. \underline{Third}, the search space for the potential optimal configuration is exponential---each operator has a distinct parallelism configuration that should be dynamically adjusted during pipeline execution, and the cluster placement of each operator’s parallel executors further affects pipeline performance by introducing cross-server multimodal data transfer overhead and synchronization costs.

As a result, existing stream processing systems may suffer severe performance degradation in multimodal data processing pipelines without including specific optimization to address the above challenges. For example, the schedulers in such systems either perform threshold-based auto-scaling on individual operators without considering end-to-end pipeline performance, or make mismatched assumptions: operators are synchronous and share homogeneous resource usage, and inter-operator data transfer overhead is negligible~\cite{dhalion,ds2,conttune,ray_data,elastic_scaling}.

We present \system, a scheduling framework that addresses these challenges through three integrated innovative components.

%\vspace{-0.5em}
\begin{itemize}[topsep=5pt, leftmargin=*]
\item
\textbf{An observation layer} that provides a noise-resilient performance model for heterogeneous multimodal pipelines. For each complex asynchronous operator in the pipeline, rather than relying on synchronous useful-time estimators that break under overlapping execution and dynamic batching, we learn a direct mapping from workload descriptors to achievable throughput using Gaussian Process regression~\cite{gpr}. To ensure the model reflects sustainable capacity rather than transient underutilization, we incorporate anomaly detection to filter observations that are dominated by upstream starvation or backpressure artifacts. This layer provides reliable capacity estimates that enable both effective scheduling and meaningful evaluation of candidate configurations during online tuning.

\item
\textbf{An adaptation layer} that performs online workload clustering and memory-constrained configuration tuning. Streaming clustering identifies distinct workload patterns without offline preprocessing. To safely optimize operator parameters under accelerator memory limits, the layer uses memory-constrained Bayesian optimization~\cite{practical_bo,mtm,bo_unknown_constraints}, treating peak accelerator memory as a black-box constraint to enable sample-efficient search while reducing OOM risk during exploration. The layer outputs candidate configurations to the scheduling layer, which decides whether and when to apply them.

\item
\textbf{A scheduling layer} that formulates a mixed-integer linear program (MILP) for joint parallelism, placement, and configuration transition under heterogeneous resource constraints. Unlike traditional autoscalers that treat workers as interchangeable, our formulation models GPUs/NPUs and CPUs as distinct resource pools with separate capacity limits. The effective source rate becomes a decision variable coupled to downstream capacity, enabling throughput maximization rather than reactive scaling. The formulation captures cluster topology and bandwidth constraints, co-locating bandwidth-intensive adjacent stages to avoid cross-node transfer bottlenecks. The MILP incorporates integer decision variables that govern rolling updates for configuration transitions, weighing the projected throughput improvement against the cold-start overhead incurred during instance restart. A migration penalty term discourages unnecessary placement changes across re-optimization rounds.
\end{itemize}

Note that these three layers form a closed control loop rather than operating independently. The capacity estimations produced by the observation layer parameterize both the MILP in the scheduling layer and the surrogate models in the adaptation layer. When the adaptation layer detects a workload regime shift and proposes a new operator configuration, the scheduling layer decides whether to adopt it by weighing the projected throughput gain against the cold-start overhead of a rolling update. Upon committing a configuration transition, the scheduling layer invalidates the observation layer's historical samples for the affected operator, triggering model re-initialization so that capacity estimates remain consistent with the active system state. This bidirectional coupling---where scheduling decisions
feed back into the observation layer and adaptation layer, and vice versa---is essential for maintaining system-wide consistency: without it, stale capacity models or uncoordinated configuration switches would erode the very throughput gains that
each layer individually provides.

We implement \system atop Ray Data and evaluate it on two production-representative
pipelines: a document processing pipeline combining PDF parsing, layout detection,
and LLM-based OCR, and a video cleaning pipeline with decoding, scene detection,
quality filtering, and captioning stages. Experiments show that Trident (\underline{\textbf{i}}) provides
robust, workload-aware capacity estimates for complex asynchronous operators,
(\underline{\textbf{ii}}) adapts operator configurations under workload regime shifts while mitigating
OOM-induced disruption, and (\underline{\textbf{iii}}) improves end-to-end scheduling decisions by
jointly optimizing parallelism, placement, and configuration transitions under fixed
heterogeneous resources. Overall, \system improves end-to-end throughput by up
to \(2.01\times\) (PDF pipeline) and \(1.88\times\) (video pipeline) over a static baseline,
with low overhead suitable for online re-optimization.

\section{Background and Motivation}
Modern data preparation pipelines increasingly process multimodal datasets that combine text documents, images, and videos, and are executed repeatedly to produce training and evaluation corpora for downstream foundation models~\cite{bai2025qwen3vltechnicalreport,opensora,data-juicer}. These pipelines typically consist of multiple stages---e.g., parsing, decoding, filtering, transformation, and model-based inference---and are often deployed using stream execution models~\cite{dataflow,flink,ray_data,spark-streaming}, where data items flow through a directed acyclic graph (DAG) of operators and intermediate results are materialized only when necessary. While streaming improves resource efficiency and reduces materialization overhead, streaming also exposes scheduling challenges that are less pronounced in traditional synchronous, record-at-a-time settings.

\subsection{Multimodal Pipeline Characteristics}
Multimodal operators exhibit highly input-dependent and non-stationary behavior. Even within a single stage, execution time and achievable throughput can vary substantially with request characteristics such as document length and layout, image resolution, or video duration and scene complexity. Many vision and video operators rely on accelerators and process data in batches; moreover, accelerator-backed inference is frequently asynchronous and uses internal dynamic/continuous batching to amortize overhead~\cite{orca-contbatch,vllm}. As a result, observed throughput can be non-linear and time-varying, and short-window measurements are often confounded by transient upstream starvation, downstream backpressure, and batching artifacts rather than reflecting sustainable operator capacity. Finally, configuration choices that improve throughput (e.g., larger batch sizes or longer sequence limits) can also increase peak device memory sharply and trigger out-of-memory failures, turning configuration selection into an online performance--safety trade-off under changing input distributions.

\subsection{Offline Resource Constraints}
In practice, multimodal data preparation pipelines are commonly executed in an offline paradigm on fixed-resource clusters of heterogeneous devices (i.e., a large number of CPUs, along with a fixed number of GPUs, NPUs, or TPUs) and under non-negligible network constraints. Operators have distinct resource affinities: some are CPU-bound, others require accelerators, and bandwidth-intensive stages can be bottlenecked by cross-node transfers of large intermediate data (e.g., frames, crops, or embeddings). Consequently, end-to-end throughput depends not only on per-operator parallelism but also on placement decisions and communication patterns. Unlike elastic cloud settings, where autoscaling can mask inefficiencies, fixed clusters require careful coordination of operator parallelism, placement, and configuration choices under hard resource budgets, while maintaining stability in the presence of workload shifts and costly operator restarts.

\section{System Overview}

\begin{figure*}[t!]
    \centering
    \includegraphics[width=0.95\textwidth]{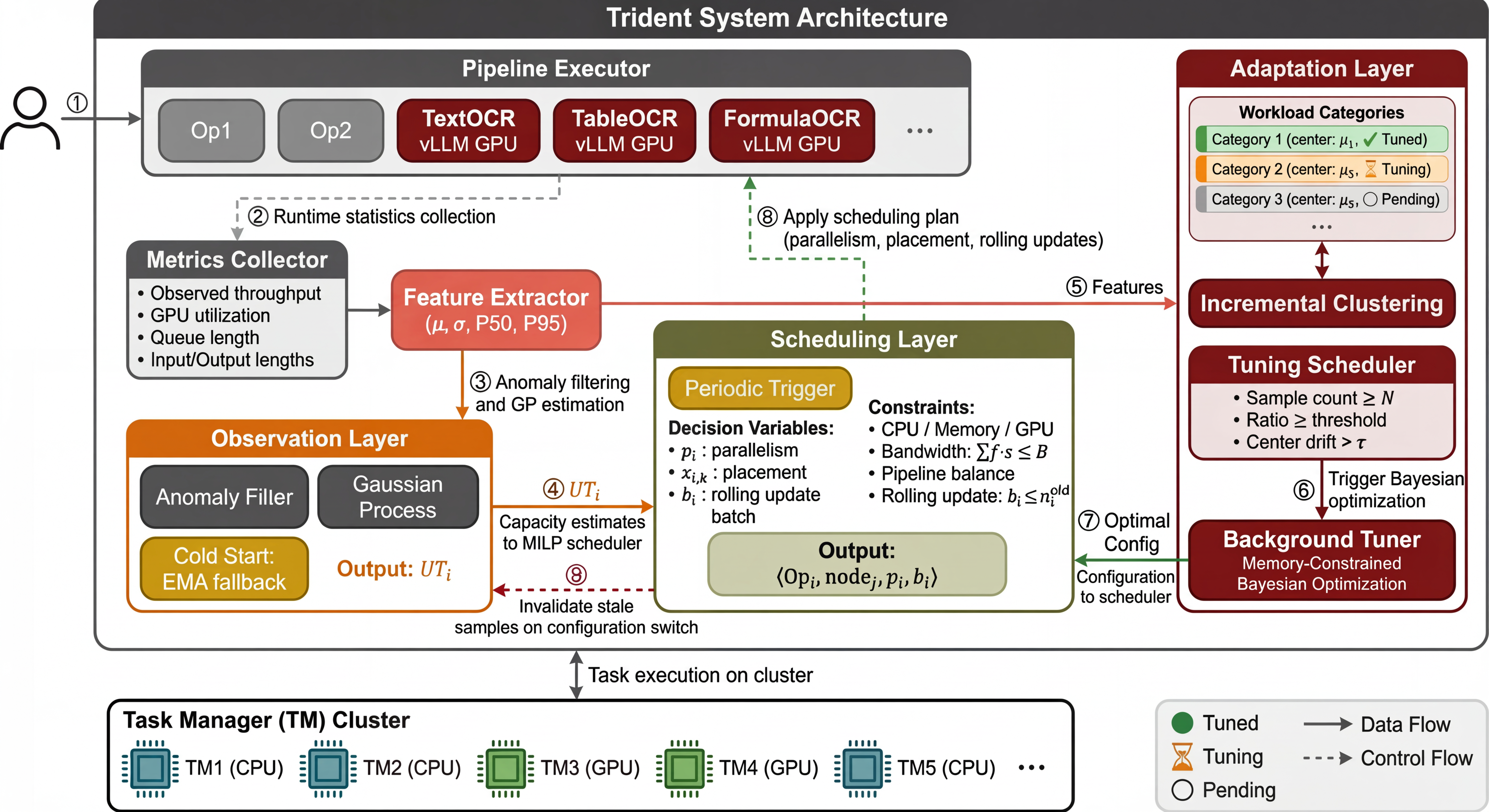}
    \caption{\system system architecture. The metrics collector gathers runtime statistics from operator instances and feeds them to two parallel paths: the observation layer filters anomalies and estimates operator throughput via Gaussian Process regression, while the adaptation layer performs workload clustering and maintains memory-aware configuration recommendations via memory-constrained Bayesian optimization. The scheduling layer integrates capacity estimates and configuration recommendations to solve an MILP that jointly optimizes parallelism, placement, and configuration transitions. Upon committing a configuration transition, the scheduling layer signals the observation layer to invalidate stale samples (path \ding{200}), ensuring capacity estimates remain consistent with the active configuration.}
    \label{fig:architecture}
    %%\vspace{-1em}
\end{figure*}

This section presents the overall architecture of \system and describes how its components interact to achieve adaptive scheduling for multimodal data preparation pipelines.

%\vspace{-0.5em}
\subsection{System Architecture}

Figure~\ref{fig:architecture} illustrates the architecture of \system, which consists of three integrated layers built atop a Ray cluster.

%\vspace{0.25em}
\noindent
\textbf{Pipeline Executor.} The pipeline executor manages the dataflow graph, instantiating operators, routing data between them, and applying parallelism updates from the scheduling layer.

%\vspace{0.25em}
\noindent
\textbf{Metrics Collector.} The metrics collector gathers runtime statistics from all operator instances---including observed throughput, resource utilization, queue lengths, per-request characteristics, and peak device memory usage---and feeds them into both the observation layer and the adaptation layer.

%\vspace{0.25em}
\noindent
\textbf{Observation Layer.} The observation layer estimates the sustainable throughput of each operator by combining anomaly filtering with Gaussian Process regression~\cite{gpr} over workload descriptors. The resulting capacity estimates feed into both the scheduling layer and the adaptation layer. Details are presented in Section~\ref{sec:observation}.

%\vspace{0.25em}
\noindent
\textbf{Adaptation Layer.} The adaptation layer tracks workload distribution shifts through online clustering and discovers optimal operator configurations via memory-constrained Bayesian optimization, which serves as a workload-aware advisor: configuration recommendations are forwarded to the scheduling layer, which decides whether and when to apply them. Details are presented in Section~\ref{sec:adaptation}.

%\vspace{0.25em}
\noindent
\textbf{Scheduling Layer.} The scheduling layer formulates a mixed-integer linear program (MILP) that jointly optimizes operator parallelism, placement, and configuration transitions under heterogeneous resource constraints. Upon committing a configuration change, the scheduling layer notifies the observation layer to invalidate stale samples, ensuring capacity estimates remain consistent with the active system state. Details are presented in Section~\ref{sec:scheduling}.

%\vspace{-0.5em}
\subsection{Control Flow}

The three layers form a closed-loop control system. The metrics collector gathers observations from operators (path \ding{193}) and feeds them into two parallel paths. The observation layer filters samples and produces capacity estimations (\ding{194}$\rightarrow$\ding{195}), while the adaptation layer performs clustering and maintains configuration recommendations (\ding{196}$\rightarrow$\ding{197}$\rightarrow$\ding{198}). The scheduling layer integrates both inputs to solve the MILP, which jointly determines parallelism, placement, and configuration transitions. Scheduling decisions are propagated back to the pipeline executor (\ding{199}). When the MILP selects a configuration change for an operator, the scheduling layer coordinates a rolling update through the pipeline executor and signals the observation layer to invalidate stale samples (\ding{200}), maintaining consistency between capacity estimates and the active system state.

\section{Observation Layer}
\label{sec:observation}

The observation layer estimates the sustainable throughput of each operator, providing capacity inputs for the scheduling layer. This section describes the challenges of throughput estimation for asynchronous operators and presents our approach combining Gaussian Process regression with anomaly detection.

%\vspace{-0.5em}
\subsection{Problem Description}
\label{sec:obs-problem}

For synchronous operators with deterministic execution, dividing processed records by useful time yields reliable capacity metrics~\cite{ds2}. However, asynchronous operators prevalent in multimodal pipelines---such as LLM inference engines with continuous batching~\cite{orca-contbatch} and batch-oriented vision models---violate this assumption: overlapping execution obscures the mapping between elapsed time and completed work, while dynamic batching makes throughput dependent on request arrival patterns rather than a fixed per-record cost. Furthermore, when an operator is starved of input or temporarily backlogged, raw measurements confound transient pipeline effects with intrinsic capacity. The observation layer, therefore, requires a throughput estimator that: (\underline{\textbf{i}}) conditions on workload characteristics to capture input-dependent performance variation, (\underline{\textbf{ii}})~quantifies prediction uncertainty for downstream decision-making, and (\underline{\textbf{iii}})~filters non-steady-state observations before they corrupt the model. We address these requirements using Gaussian Process regression~\cite{gpr} in combination with a two-stage anomaly filtering pipeline.

\subsection{Throughput Modeling by Gaussian Process}

We model the relationship between workload characteristics and operator throughput using Gaussian Process (GP) regression~\cite{gpr}. GP is well suited to this setting for two reasons: (\underline{\textbf{i}}) flexible nonlinear modeling without a fixed functional form, and (\underline{\textbf{ii}}) explicit uncertainty quantification, which is valuable for both extrapolation to unseen workload regions and model-based anomaly detection.

%\vspace{0.25em}
\noindent
\textbf{Feature Representation.} For each asynchronous operator, we define a feature vector $\mathbf{x}$ capturing workload characteristics relevant to its throughput. For LLM inference operators, features include the mean and standard deviation of input token lengths ($\mu_{\text{in}}$, $\sigma_{\text{in}}$) and output token lengths ($\mu_{\text{out}}$, $\sigma_{\text{out}}$). Specifically, $\mathbf{x} = (\mu_{\text{in}}, \sigma_{\text{in}}, \mu_{\text{out}}, \sigma_{\text{out}})$. For vision operators, features include image resolution and batch size. The specific features are operator-dependent and configured at pipeline definition time.

%\vspace{0.25em}
\noindent
\textbf{Model Formulation.} We model throughput $y$ as a function of workload features formulated below:
\begin{equation}
y = f(\mathbf{x}) + \epsilon, \quad f \sim \mathcal{GP}(m(\mathbf{x}), k(\mathbf{x}, \mathbf{x}'))
\end{equation}
where $m(\mathbf{x})$ is the mean function, $k(\mathbf{x}, \mathbf{x}')$ is the covariance kernel, and $\epsilon \sim \mathcal{N}(0, \sigma_n^2)$ represents observation noise. We use a constant mean function and the Mat\'{e}rn 5/2 kernel, which provides smoothness appropriate for physical throughput relationships while remaining robust to local variations.

%\vspace{0.25em}
\noindent
\textbf{Incremental Updates.} As new filtered observations arrive, we update the GP model incrementally. To bound computational cost, we maintain a fixed-size inducing point set using a sliding window over recent observations. When the observation buffer exceeds capacity, we remove the oldest samples while preserving coverage across the observed feature space.

%\vspace{0.25em}
\noindent
\textbf{Throughput Prediction.} Given the current workload characteristics $\mathbf{x}^*$, the GP provides a predictive distribution:
\begin{equation}
p(y^*|\mathbf{x}^*, \mathcal{D}) = \mathcal{N}(\mu^*, \sigma^{*2})
\end{equation}
where $\mathcal{D}$ denotes the training set of filtered observations. We report the predictive mean $\mu^*$ as the capacity estimate to the scheduling layer. 
The predictive variance \(\sigma^{*2}\) quantifies estimation confidence and supports two use cases: (i) identifying predictions in unexplored workload regions that require caution and (ii) enabling model-based anomaly detection (Section~\ref{sec:anomaly}).

%\vspace{-0.5em}
\subsection{Anomaly Detection}
\label{sec:anomaly}

Not all throughput observations reflect sustainable operator capacity. We apply a two-stage filtering pipeline to identify and discard anomalous samples before they are included in the GP training set.

%\vspace{0.25em}
\noindent
\textbf{Stage 1: Signal-based Filtering.} The first stage uses lightweight, readily available runtime signals to detect and discard observations collected under non-steady-state conditions.

\begin{itemize}[topsep=5pt, leftmargin=*]
\item
\textit{Utilization-based filtering.} We monitor GPU/NPU/TPU utilization for accelerated operators and CPU utilization for compute-bound stages. Observations collected when utilization falls below a threshold $\tau_u$ are excluded, preventing underestimation caused by upstream starvation.

\item\textit{Queue-based filtering.} We track queue length over sliding windows and detect two anomalous patterns: draining queues indicate the operator is outpacing its input supply, while rapidly growing queues suggest a temporary backlog that inflates apparent throughput. Thus, observations during these transient phases will be excluded.
\end{itemize}

%\vspace{0.25em}
\noindent
\textbf{Stage 2: Model-based Filtering.} Samples passing the first stage may still contain outliers due to transient system effects (e.g., garbage collection pauses or resource contention). For each candidate observation $(\mathbf{x}, y)$, we query the current GP model for a predictive distribution at $\mathbf{x}$:
\begin{equation}
p(y^*|\mathbf{x}, \mathcal{D}) = \mathcal{N}(\mu, \sigma^2)
\end{equation}
We compute the standardized residual $z = (y - \mu)/\sigma$ and reject observations where $|z| > \tau_z$, preventing outliers from corrupting subsequent updates.

We want to emphasize that the two stages are complementary: the signal-based filtering catches systematic issues detectable from runtime metrics, while model-based filtering catches sporadic anomalies that escape the first stage.

\subsection{Cold Start and Sample Invalidation}
\label{sec:cold-start}

During system startup or after a configuration change, insufficient valid samples may be available for reliable GP prediction. When the number of filtered observations falls below a threshold~$n_{\min}$, we bypass the GP model and report an exponential moving average (EMA) of recent throughput observations as the capacity estimate, with only signal-based filtering (Stage~1) active. Once the observation count exceeds~$n_{\min}$, we transition to GP-based prediction with the full two-stage filtering pipeline.
When the scheduling layer commits a configuration transition for an operator, historical samples collected under the previous configuration no longer reflect the operator's behavior. Upon receiving the transition notification, the observation layer clears its observation buffer and resets the GP model for the affected operator, triggering a return to EMA-based estimation until sufficient new samples accumulate.

\section{Adaptation Layer}
\label{sec:adaptation}

The adaptation layer detects workload distribution shifts and identifies near-optimal operator configurations for each workload pattern. As a workload-aware advisor, the adaptation layer performs online workload categorization and configuration tuning and forwards recommendations to the scheduling layer, which decides whether and when to apply configuration transitions as part of its unified MILP optimization. Algorithm~\ref{alg:adaptation} presents the overall control flow.

%\vspace{-0.5em}
\subsection{Problem Description}
\label{sec:adapt-problem}

Operator configurations such as batch size, maximum sequence length, and tiling strategy significantly affect both throughput and peak accelerator memory. Optimal settings are workload-dependent: shorter inputs favor larger batches to improve device utilization, whereas longer inputs require smaller batches to avoid memory exhaustion. The exact memory footprint depends on framework allocators, kernel fusion, and caching behavior and therefore behaves as a black-box function of the configuration and input distribution. Naively exploring the configuration space risks frequent out-of-memory (OOM) failures that cause worker restarts and prolonged throughput drops. The adaptation layer must therefore (i)~detect when the workload distribution has shifted sufficiently to warrant re-tuning, and (ii)~search for high-throughput configurations sample-efficiently while bounding OOM risk. We achieve this through online workload clustering coupled with memory-constrained Bayesian optimization.

\begin{algorithm}[t]
\caption{Adaptation Layer Control Flow}
\label{alg:adaptation}
\begin{algorithmic}[1]
\Require Feature vector $\mathbf{x}$ for incoming request
\Ensure Updated cluster state and configuration recommendations

\State \textcolor{gray}{\textit{// Phase 1: Workload Categorization}}
\State $c \gets \textsc{AssignCluster}(\mathbf{x})$
\State $\textsc{UpdateClusterStats}(c, \mathbf{x})$

\State \textcolor{gray}{\textit{// Phase 2: Trigger Condition Check}}
\If{$\textsc{ShouldTriggerTuning}(c)$}
    \State $\textsc{EnqueueTuningJob}(c)$
\EndIf

\State \textcolor{gray}{\textit{// Phase 3: Forward Recommendations}}
\State $\textsc{UpdateMatchHistory}(c)$
\State $c_{dom} \gets \textsc{GetDominantCluster}()$
\If{$c_{dom}.s = \textit{Tuned}$}
    \State $\textsc{ForwardRecommendation}(c_{dom}.\text{config}, c_{dom}.UT^{cand})$
\EndIf
\end{algorithmic}
\end{algorithm}

%\vspace{-0.5em}
\subsection{Workload Categorization}
\label{sec:workload-cat}

We maintain a set of workload categories using an online clustering algorithm that updates incrementally as new samples arrive.

%\vspace{0.25em}
\noindent
\textbf{Cluster Representation.} Each cluster $C_i$ is represented as a tuple $(\boldsymbol{\mu}_i, N_i, s_i, \mathbf{\theta}_i^*)$, where $\boldsymbol{\mu}_i$ denotes the centroid, $N_i$ the sample count, $s_i \in \{\text{Pending}, \text{Tuning}, \text{Tuned}\}$ the tuning status, and $\mathbf{\theta}_i^*$ the optimal configuration if available. We limit the number of clusters to $L_{\max}$ to bound memory usage.

%\vspace{0.25em}
\noindent
\textbf{Feature Extraction.} For each processed request, we extract a low-dimensional feature vector $\mathbf{x}$ characterizing the workload. Features are operator-specific: for LLM operators, we use input token length and output token length; for vision operators, we use image resolution and aspect ratio. 

%\vspace{0.25em}
\noindent
\textbf{Cluster Assignment and Update.} When a new sample arrives,
we compute its distance to each existing centroid. If the minimum
distance falls below the threshold $\tau_d$, we assign the sample to
the nearest cluster and update the centroid incrementally via
$\boldsymbol{\mu}_{\text{nearest}} \leftarrow \boldsymbol{\mu}_{\text{nearest}}
+ \frac{1}{N_{\text{nearest}}+1}(\mathbf{x} - \boldsymbol{\mu}_{\text{nearest}})$.
Otherwise, we create a new cluster with the sample as its centroid.
If the number of clusters reaches the limit $L_{\max}$, the two
closest clusters are merged before the new one is added.

%\vspace{0.25em}
\noindent
\textbf{Cluster Maintenance.} To adapt to shifting distributions, we periodically apply exponential decay to sample counts: $N_i \leftarrow \gamma \cdot N_i$ where $\gamma < 1$. Clusters whose counts fall below a minimum threshold are removed, allowing the system to forget obsolete patterns.

%\vspace{-0.5em}
\subsection{Configuration Optimization}
\label{sec:config-opt}

\system optimizes operator-level runtime configurations online to maximize sustainable throughput under the current dominant workload cluster. For accelerator-backed operators, aggressive exploration of configuration parameters (e.g., batch size, maximum sequence length, tile size) can easily trigger device out-of-memory (OOM) failures, causing worker restarts and prolonged throughput drops. Peak device memory is difficult to model analytically because peak memory depends on framework allocators, kernel fusion, and caching behavior. \system therefore treats device memory as a black-box constraint and uses memory-constrained Bayesian optimization during tuning so that recommended configurations are both high-throughput and OOM-safe, without requiring additional mechanisms in the scheduling layer.

Formally, for a tunable operator $i$ and dominant workload cluster $c$ described by a feature vector $\mathbf{x}_c$, \system chooses a configuration $\theta \in \Theta_i$ by solving:
\begin{equation}
\max_{\theta \in \Theta_i}\ UT_i(\theta; \mathbf{x}_c)
\quad \text{s.t.}\quad
Mem_i(\theta; \mathbf{x}_c) \le M_i^{cap} - \Delta_i ,
\label{eq:mem_constrained_opt}
\end{equation}
where $UT_i(\theta; \mathbf{x}_c)$ is sustainable throughput, $Mem_i(\theta; \mathbf{x}_c)$ is peak device memory during execution, $M_i^{cap}$ is the available device memory capacity for the operator instance, and $\Delta_i$ is a safety margin that accounts for allocator fragmentation, runtime caching, and measurement noise.
Because both throughput and peak memory are expensive to evaluate and do not admit closed-form expressions, \system maintains two probabilistic surrogate models learned from online measurements (Gaussian Processes in our implementation):
\begin{align}
UT_i(\theta; \mathbf{x}_c) &\sim \mathcal{N}\!\left(\mu_{UT}(\theta; \mathbf{x}_c),\ \sigma^2_{UT}(\theta; \mathbf{x}_c)\right),
\label{eq:gp_ut}
\\
Mem_i(\theta; \mathbf{x}_c) &\sim \mathcal{N}\!\left(\mu_{M}(\theta; \mathbf{x}_c),\ \sigma^2_{M}(\theta; \mathbf{x}_c)\right).
\label{eq:gp_mem}
\end{align}
Each tuning evaluation records $(UT_i(\theta; \mathbf{x}_c), Mem_i(\theta; \mathbf{x}_c))$. If an evaluation triggers OOM, \system marks $\theta$ as infeasible so subsequent proposals avoid repeatedly exploring unsafe regions.
From the memory surrogate in Eq.~\eqref{eq:gp_mem}, \system computes the probability that $\theta$ satisfies the memory constraint:
\begin{equation}
PoF(\theta)= \Pr\!\left(Mem_i(\theta; \mathbf{x}_c) \le M_i^{cap}-\Delta_i\right)= \Phi\!\left(\frac{M_i^{cap}-\Delta_i-\mu_{M}(\theta; \mathbf{x}_c)}{\sigma_{M}(\theta; \mathbf{x}_c)}\right)
\label{eq:pof}
\end{equation}
where $\Phi(\cdot)$ is the standard normal CDF.

\noindent Let $UT_i^{+}$ denote the best observed sustainable throughput among feasible (non-OOM) configurations. \system uses expected improvement on throughput, denoted $EI_{UT}(\theta)$, and combines $EI_{UT}(\theta)$ with feasibility to define a constrained acquisition function:
\begin{equation}
\alpha(\theta) = EI_{UT}(\theta)\cdot PoF(\theta).
\label{eq:cei}
\end{equation}
The next configuration is selected by maximizing $\alpha(\theta)$ while enforcing a minimum feasibility threshold:
\begin{equation}
\theta_{t+1} = \arg\max_{\theta\in\Theta_i}\ \alpha(\theta)
\quad \text{s.t.}\quad
PoF(\theta) \ge \eta ,
\label{eq:pof_gate}
\end{equation}
where $\eta$ controls the exploration--safety trade-off. After exhausting the tuning budget, \system recommends the best configuration candidate with the highest predicted throughput among candidates satisfying $PoF(\theta)\ge \eta$, keeping the safety mechanism entirely within the tuning loop.

% \section{Scheduling Layer: Joint Parallelism and Placement Optimization}
% \subsection{System Model}
% \subsection{MILP Formulation}
% \subsection{Handling Bandwidth Constraints}
% \subsection{Migration Cost and Stability}

%\vspace{-0.5em}
\section{Scheduling Layer}
\label{sec:scheduling}

The scheduling layer determines operator parallelism and placement to maximize pipeline throughput under fixed resource constraints. This section formulates the scheduling problem as a mixed-integer linear program (MILP) and describes the solution approach.

%\vspace{-0.5em}
\subsection{Problem Description}

Given a pipeline of $n$ operators processing data through a cluster of $K$ heterogeneous nodes, the scheduling layer must decide how many instances of each operator to run and where to place them. This problem differs from traditional stream processing autoscaling in three respects.
\underline{First}, resources are fixed rather than elastic. Offline data preparation runs on dedicated clusters where CPU cores, memory, and accelerator devices cannot be provisioned on demand. Unlike online streaming settings where the source rate is externally imposed by incoming data streams, in offline batch processing, the source rate is itself a decision variable: we seek to maximize end-to-end throughput subject to hard capacity limits on each node.
\underline{Second}, operator placement matters. When adjacent operators reside on different nodes, intermediate data must traverse the network. For bandwidth-intensive stages---such as those following decompression or frame extraction---network capacity can become the binding constraint. Placement-unaware scheduling misses opportunities to co-locate communicating operators and reduce cross-node traffic.
\underline{Third}, operator configuration selection interacts with parallelism and placement. Switching to a better-matched configuration promises higher steady-state throughput, but each transition incurs cold-start overhead from instance restarts and observation warm-up. The scheduling layer must therefore weigh the projected gain against the transition cost and decide both \emph{whether} and \emph{how fast} to apply configuration changes via rolling updates.
Our scheduling layer addresses these challenges through an MILP formulation that jointly optimizes parallelism, placement, and configuration transitions while modeling heterogeneous resources, data amplification, network bandwidth, and cold-start overhead.

%\vspace{-0.5em}
\subsection{System Model}

\noindent
\textbf{Cluster Resources.}
The cluster consists of $K$ nodes, each with heterogeneous resource capacities. Node $k$ provides $C_k^{cpu}$ CPU cores, $C_k^{mem}$ GB of memory, and $C_k^{gpu}$ GPU devices. These capacities are fixed and known at scheduling time.

%\vspace{0.25em}
\noindent
\textbf{Pipeline Structure.} The pipeline comprises $n$ operators arranged in a linear dataflow: operator 1 receives raw input, each operator $i$ passes its output to operator $i+1$, and operator $n$ produces final results. Each operator $i$ has the following characteristics, obtained from the observation layer and adaptation layer:

\begin{itemize}[topsep=5pt, leftmargin=*]
    \item \textit{Unit throughput} $UT_i$: the processing rate of a single instance, in records per second. This value is provided by the observation layer's Gaussian Process model based on current workload characteristics.
    
    \item \textit{Resource requirements}: CPU cores $u_i$, memory $m_i$ GB, and GPU devices $g_i$ consumed by each instance.
    
    \item \textit{Output data size} $d^{out}_i$: the size in MB of each record produced by operator $i$.
    
    \item \textit{Data amplification factor} $D_i$: the ratio of operator $i$'s input volume to the original pipeline input. This captures how data expands or contracts through the pipeline. For example, if operator 1 splits each input document into 10 pages, then $D_2 = 10$; if operator 2 filters half, then $D_3 = 5$.
\end{itemize}

Let $D_o$ denote the amplification factor at the pipeline output. The relationship between operator $i$'s local throughput and the equivalent pipeline input rate is governed by $D_i$: processing $D_i$ records at operator $i$ corresponds to one record of original input.

%\vspace{0.25em}
\noindent
\textbf{Current Deployment State.} For online rescheduling, we track the current deployment: $\bar{x}_{i,k}$ denotes the number of instances of operator $i$ currently running on node $k$. At initial startup, $\bar{x}_{i,k} = 0$ for all $i, k$. Each operator also has associated migration costs: $h_i^{start}$ seconds to launch a new instance and $h_i^{stop}$ seconds to terminate an existing one.

%\vspace{0.25em}
\noindent
\textbf{Configuration State.} For each tunable operator $i$, the system maintains the current rolling update state: $n_i^{old}$ instances running the current configuration and $n_i^{new}$ instances already transitioned to the candidate configuration from previous scheduling rounds. At initial startup or when no rolling update is in progress, $n_i^{old} = p_i$ and $n_i^{new} = 0$.

The adaptation layer provides the candidate configuration's estimated unit throughput $UT_i^{cand}$, while the observation layer provides the current configuration's measured unit throughput $UT_i^{cur}$. The cold-start overhead per instance is $h_i^{cold}$, encompassing both the instance restart time and the observation warm-up period during which the GP model is unavailable.

%\vspace{-0.5em}
\subsection{Decision Variables}

The MILP formulation employs the following decision variables:

\begin{itemize}[topsep=5pt, leftmargin=*]
    \item $p_i \in \mathbb{Z}^+$: total number of instances (parallelism) for operator $i$.

    \item $x_{i,k} \in \mathbb{Z}^{\geq 0}$: number of instances of operator $i$ placed on node $k$.

    \item $b_i \in \mathbb{Z}^{\geq 0}$: number of instances of operator $i$ to transition from current to candidate configuration in this scheduling round (rolling update batch size). For non-tunable operators or those without a ready candidate configuration ($s_i \neq \textit{Tuned}$), $b_i = 0$.

    \item $w_{i,k,l} \in \mathbb{Z}^{\geq 0}$: data flow units from operator $i$ on node $k$ to operator $i+1$ on node $l$. Each unit represents one instance's worth of output capacity.
    
    \item $\delta_{i,k}^+, \delta_{i,k}^- \in \mathbb{Z}^{\geq 0}$: instances of operator $i$ added to or removed from node $k$ relative to current deployment.
    
    \item $T \geq 0$: pipeline throughput in original input records per second.
    
    \item $E_{max} \geq 0$: maximum egress traffic (MB/s) across all nodes.
    
    \item $J_{mig} \geq 0$: total migration cost in seconds.
\end{itemize}

%\vspace{-0.5em}
\subsection{Optimization Objective}

The primary goal is maximizing throughput $T$. Secondary objectives include minimizing peak network load to avoid bandwidth bottlenecks and minimizing migration cost to maintain deployment stability. We combine these through weighted summation:

\begin{equation}
\max \quad T - \lambda_1 \cdot E_{max} - \lambda_2 \cdot J_{mig}
\label{eq:objective}
\end{equation}

\noindent The coefficients $\lambda_1$ and $\lambda_2$ are small positive values that establish a priority ordering: throughput dominates, with network balance and migration cost serving as tiebreakers. In practice, we set $\lambda_1 = 10^{-4}$ and $\lambda_2 = 10^{-6}$ to ensure that the secondary objectives influence the solution only when throughput is equal.

%\vspace{-0.5em}
\subsection{Constraints}

\noindent\textbf{Throughput Constraints.} System throughput cannot exceed any operator's processing capacity. When an operator undergoes a rolling update ($b_i > 0$), its instances contribute at different rates depending on their configuration state.
At the start of a scheduling round, operator $i$ has $n_i^{new}$ instances already on the candidate configuration and $n_i^{old}$ on the current configuration. The MILP decides total parallelism $p_i$ and rolling batch size $b_i$. Since already-transitioned instances are never rolled back, $p_i \geq n_i^{new}$ holds by design. Of the $p_i$ instances in the new deployment:

\begin{itemize}[topsep=3pt, leftmargin=*]
    \item $n_i^{new}$ instances continue on the candidate configuration at rate $UT_i^{cand}$;
    \item $b_i$ instances are transitioning: each is restarted and incurs cold-start overhead $h_i^{cold}$, yielding effective rate $\hat{UT}_i$ (Eq.~\eqref{eq:effective_throughput});
    \item the remaining $p_i - n_i^{new} - b_i$ instances stay on the current configuration at rate $UT_i^{cur}$.
\end{itemize}

The effective throughput of a transitioning instance accounts for the cold-start overhead within the scheduling window:

\begin{equation}
\hat{UT}_i = UT_i^{cand} \cdot \max\left(0,\; 1 - \frac{h_i^{cold}}{T_{sched}}\right)
\label{eq:effective_throughput}
\end{equation}

\noindent Since $h_i^{cold}$ and $T_{sched}$ are known constants, $\hat{UT}_i$ is precomputed before MILP construction. Therefore, $\hat{UT}_i$ does not introduce nonlinearity. 
We define:
\begin{equation}
p_i^{stay} = p_i - n_i^{new} - b_i
\label{eq:start}
\end{equation}

\noindent For the $i$-th operator, the throughput constraint becomes:

\begin{equation}
T \leq \frac{D_o}{D_i} \cdot \left[ p_i^{stay} \cdot UT_i^{cur} + n_i^{new} \cdot UT_i^{cand} + b_i \cdot \hat{UT}_i \right]%, \ \forall i \in \{1, \ldots, n\}
\end{equation}

\noindent Since $n_i^{new}$ is a constant (determined by the previous round's state), all terms are linear in the decision variables $p_i$ and $b_i$. When no configuration transition is active or scheduled ($b_i = 0$ and $n_i^{new} = 0$), this reduces to the standard constraint $T \leq \frac{D_o}{D_i} \cdot p_i \cdot UT_i^{cur}$.

%\vspace{0.25em}
\noindent 
\textbf{Placement Consistency.}
The sum of instances across all nodes must equal total parallelism:

\begin{equation}
\sum_{k=1}^{K} x_{i,k} = p_i, \quad \forall i \in \{1, \ldots, n\}
\end{equation}

%\vspace{0.25em}
\noindent 
\textbf{Resource Capacity Constraints.} Each node's resource consumption must not exceed its capacity:

\begin{align}
\sum_{i=1}^{n} u_i \cdot x_{i,k} &\leq C_k^{cpu}, \quad \forall k \in \{1, \ldots, K\} \\
\sum_{i=1}^{n} m_i \cdot x_{i,k} &\leq C_k^{mem}, \quad \forall k \in \{1, \ldots, K\} \\
\sum_{i=1}^{n} g_i \cdot x_{i,k} &\leq C_k^{gpu}, \quad \forall k \in \{1, \ldots, K\}
\end{align}

\noindent These constraints model heterogeneous resources as distinct pools, unlike traditional autoscalers that treat all workers as interchangeable, uniform capacity.

%\vspace{0.25em}
\noindent 
\textbf{Data Flow Conservation.} Data produced by operator $i$ on node $k$ must be consumed by operator $i+1$ instances somewhere in the cluster. The flow variables $w_{i,k,l}$ model this routing:

\begin{align}
\sum_{l=1}^{K} w_{i,k,l} &= x_{i,k}, \quad \forall i \in \{1, \ldots, n-1\}, \forall k \\
\sum_{k=1}^{K} w_{i,k,l} &= x_{i+1,l}, \quad \forall i \in \{1, \ldots, n-1\}, \forall l
\end{align}

\noindent The first constraint ensures each instance of operator $i$ on node $k$ sends its output somewhere; the second ensures each instance of operator $i+1$ on node $l$ receives input from somewhere. Together, these enforce flow balance across the operator boundary.

%\vspace{0.25em}
\noindent 
\textbf{Network Bandwidth Constraints.} Cross-node data transfer consumes network bandwidth. For operator $i$ on node $k$, the egress traffic to other nodes is $(x_{i,k} - w_{i,k,k}) \cdot UT_i \cdot d^{out}_i$ MB/s, where $w_{i,k,k}$ represents local transfers that bypass the network. Total egress from node $k$ is bounded by $E_{max}$:

\begin{equation}
\sum_{i=1}^{n-1} (x_{i,k} - w_{i,k,k}) \cdot UT_i \cdot d^{out}_i \leq E_{max}, \quad \forall k \in \{1, \ldots, K\}
\end{equation}

\noindent Minimizing $E_{max}$ in the objective encourages co-location of adjacent operators, reducing cross-node traffic.

%\vspace{0.25em}
\noindent 
\textbf{Migration Accounting.} The difference between the new and current deployment determines migration:

\begin{equation}
x_{i,k} = \bar{x}_{i,k} + \delta_{i,k}^+ - \delta_{i,k}^-, \quad \forall i, k
\end{equation}

\noindent Total migration cost aggregates startup and shutdown overhead:

\begin{equation}
J_{mig} = \sum_{i=1}^{n} \sum_{k=1}^{K} \left( h_i^{start} \cdot \delta_{i,k}^+ + h_i^{stop} \cdot \delta_{i,k}^- \right)
\end{equation}

\noindent Including $J_{mig}$ in the objective with coefficient $\lambda_2$ encourages the solver to prefer solutions close to the current deployment when throughput is similar, reducing unnecessary churn.

%\vspace{0.25em}
\noindent 
\textbf{Rolling Update Constraints.}
The rolling update process is governed by four constraints:
\begin{align}
& p_i \geq n_i^{new}, && \forall i \label{eq:no_rollback} \\
& b_i \leq n_i^{old}, && \forall i \\
& b_i \leq B_i^{max}, && \forall i \\
& p_i^{stay} = p_i - n_i^{new} - b_i \geq 0, && \forall i
\label{eq:end}
\end{align}

\noindent Constraint~\eqref{eq:no_rollback} encodes the no-rollback semantics discussed above: since already-transitioned instances remain on the candidate configuration, the solver must allocate at least $n_i^{new}$ instances. This simultaneously eliminates the $\min(n_i^{new}, p_i)$ nonlinearity from the throughput constraint. The second and third bound the batch size by the number of remaining old instances and a configurable maximum, respectively. The fourth ensures non-negativity of the staying instances. The MILP selects $b_i > 0$ only when $\hat{UT}_i > UT_i^{cur}$, i.e., when the candidate configuration's throughput, discounted by cold-start overhead, exceeds the current configuration's throughput. When $UT_i^{cand}$ is only marginally better or the cold-start penalty is large relative to $T_{sched}$, the solver defers the transition, avoiding unnecessary disruption.
To avoid configuration thrashing, the scheduling layer enforces a single-transition invariant: each operator may have at most one pending configuration transition at any time. If operator $i$ is mid-transition ($n_i^{old} > 0$), new recommendations from the adaptation layer are buffered but not acted upon until the current transition completes ($n_i^{old} = 0$). This ensures the formulation only needs to track two configuration states per operator (current and candidate), keeping the MILP tractable.
Combining the objective (Eq.~\eqref{eq:objective}) with constraints (Eqs.~\eqref{eq:start}--\eqref{eq:end}) yields the complete MILP formulation, which we solve as described next.

\subsection{Periodic Rescheduling}

Algorithm~\ref{alg:scheduling} describes the periodic rescheduling loop.
Before each round, the scheduling layer queries the observation layer
for current capacity estimates $\{UT^{cur}_i\}$ and the adaptation layer
for candidate configurations and their tuning status.
For each tunable operator with a ready candidate ($s_i = \text{Tuned}$),
the cold-start discounted throughput $\hat{UT}_i$ is precomputed via
Eq.~\eqref{eq:effective_throughput}.
These values parameterize the MILP, which is solved by a dedicated Ray actor running asynchronously;
the scheduler continues operating under the most recent feasible
solution until an updated plan is returned.

\begin{algorithm}[t]
\caption{Periodic Rescheduling}
\label{alg:scheduling}
\begin{algorithmic}[1]
\Require Rescheduling interval $T_{sched}$, MILP solver, current state
  $\{\bar{x}_{i,k}\}$, rolling update state $\{n^{old}_i, n^{new}_i\}$
\Loop
  \State $\{UT^{cur}_i\} \gets$ query Observation Layer
  \State $\{UT^{cand}_i, s_i\} \gets$ query Adaptation Layer
  \For{each tunable operator $i$ with $s_i = \text{Tuned}$}
    \State $\hat{UT}_i \gets UT^{cand}_i \cdot \max(0,\; 1 - h^{cold}_i / T_{sched})$
  \EndFor
  \State Construct and solve MILP $\;\to\; (p^*, x^*, b^*, w^*)$
  \State Apply placement and configuration changes 
  \State $\bar{x}_{i,k} \gets x^*_{i,k}$ for all $i, k$
  \State \textbf{sleep} $T_{sched}$
\EndLoop
\end{algorithmic}
\end{algorithm}

The solver output is applied incrementally: only the delta between the
current and target placements is executed, launching or terminating
instances gracefully as needed to minimize disruption to in-flight processing.
For operators with $b^*_i > 0$, the designated instances are restarted
with the candidate configuration via rolling update, and the rolling
update state is updated accordingly ($n^{new}_i \mathrel{+}= b^*_i$,
$n^{old}_i \mathrel{-}= b^*_i$).
Upon committing a configuration transition, the scheduling layer
signals the observation layer to invalidate historical samples for
the affected operator, triggering a return to EMA-based estimation
(Section~\ref{sec:cold-start}) until new samples accumulate under
the updated configuration.

\section{Implementation}
\label{sec:implementation}

We implement \textsc{\system} as an extension to Ray Data v2.46.0~\cite{ray_data}, a distributed data processing library built on the Ray framework~\cite{ray}.

For the observation layer, we extend Ray Data's metrics infrastructure to capture per-operator timestamps, record counts, and queue statistics. Workers maintain local buffers that flush to a central coordinator at configurable intervals. The coordinator maintains sliding-window statistics and tracks data amplification factors using exponential smoothing. The GP regression module is embedded in the scheduler process using a lightweight implementation that updates incrementally.

The adaptation layer maintains per-operator Bayesian optimization state, including surrogate models for both throughput and peak device memory. Peak device memory is collected from device telemetry (e.g., NVML on GPUs or vendor runtime APIs on NPUs), and OOM events are detected from worker exceptions and restart reasons; these signals update the feasibility model used by the constrained acquisition function.

For the scheduling layer, we extend Ray Data's abstract autoscaler interface with MILP-based optimization. To avoid blocking the scheduling loop, MILP solving is delegated to a dedicated Ray actor; the autoscaler continues operating under the most recent feasible solution until an updated plan is returned. A key modification is the scale-down policy: Ray Data's default policy terminates only idle instances, whereas our implementation marks busy instances for termination when reduced parallelism is indicated. Marked instances complete current tasks before releasing resources, enabling prompt scaling without task preemption.

% \section{Evaluation}
% \subsection{Experimental Setup}
% \subsection{End-to-End Performance}
% \subsection{Observation Layer Accuracy}
% \subsection{Adaptation Layer Effectiveness}
% \subsection{Scheduling Layer Analysis}
% \subsection{Ablation Study}
\section{EVALUATION}

We evaluate \system to answer the following research questions:

\begin{itemize}[topsep=5pt, leftmargin=*]
  \item \textbf{RQ1}: \textit{How does \system perform compared to existing scheduling approaches in end-to-end throughput?}
  \item \textbf{RQ2}: \textit{How does \system's \emph{scheduling layer} perform compared to existing schedulers under identical observation/adaptation support (i.e., a controlled comparison)?}
  \item \textbf{RQ3}: \textit{How accurate is the throughput estimation in the observation layer?}
  \item \textbf{RQ4}: \textit{Can the adaptation layer effectively detect workload shifts, discover high-throughput configurations, and maintain memory safety during online exploration?}
  \item \textbf{RQ5}: \textit{How does each component contribute to overall performance?}
  \item \textbf{RQ6}: Is the system overhead acceptable?
\end{itemize}

%\vspace{-0.5em}
\subsection{Experimental Setup}
\label{sec:exp_setup}

\textbf{Cluster Configuration.}
We conduct experiments on a cluster of 8 servers, each equipped with 8 Huawei Ascend 910B NPUs, 256 CPU cores, and 1\,TB memory, interconnected via 100\,Gbps network.

%\vspace{0.25em}
\noindent
\textbf{Datasets.}
We use two datasets spanning distinct modalities.
The \emph{PDF dataset} contains \(\sim 200k\) documents of three types---academic papers, annual reports, and financial reports---processed sequentially by type.
The \emph{video dataset} contains \(\sim 410k\) clips in two categories: short-form (10--30\,s, \(\leq\)720p) and long-form (5--10\,min, 1080p--4K), likewise processed sequentially by category.

%\vspace{0.25em}
\noindent
\textbf{Pipeline Configurations.}
The \emph{PDF curation pipeline} comprises 17 operators across five stages (file I/O, parsing and layout detection, block segmentation, modality-specific OCR, and aggregation), expanding each document into ${\sim}120$ content blocks. Three LLM-based OCR operators each require 1 NPU; the remaining operators are CPU-bound.
The \emph{video curation pipeline} comprises 9 operators across four stages (scene-based splitting, aesthetic filtering, OCR-based text filtering, and LLM-based captioning). Three NPU operators---CLIP-based scoring, CRAFT text detection, and Qwen2.5-VL-7B captioning---each require 1 NPU, with the remaining six being CPU-bound.

%\vspace{0.25em}
\noindent
\textbf{Baselines.}
We compare \system with Static, Ray Data's default autoscaler~\cite{ray_data}, DS2~\cite{ds2}, ContTune~\cite{conttune}, and SCOOT~\cite{scoot}; baseline details and adaptations are described in Section~\ref{sec:eval:e2e} and Section~\ref{sec:eval:scheduling}.

\subsection{End-to-End System Comparison (RQ1)}
\label{sec:eval:e2e}

\begin{figure}[t]
  \centering
  \includegraphics[width=0.6\linewidth]{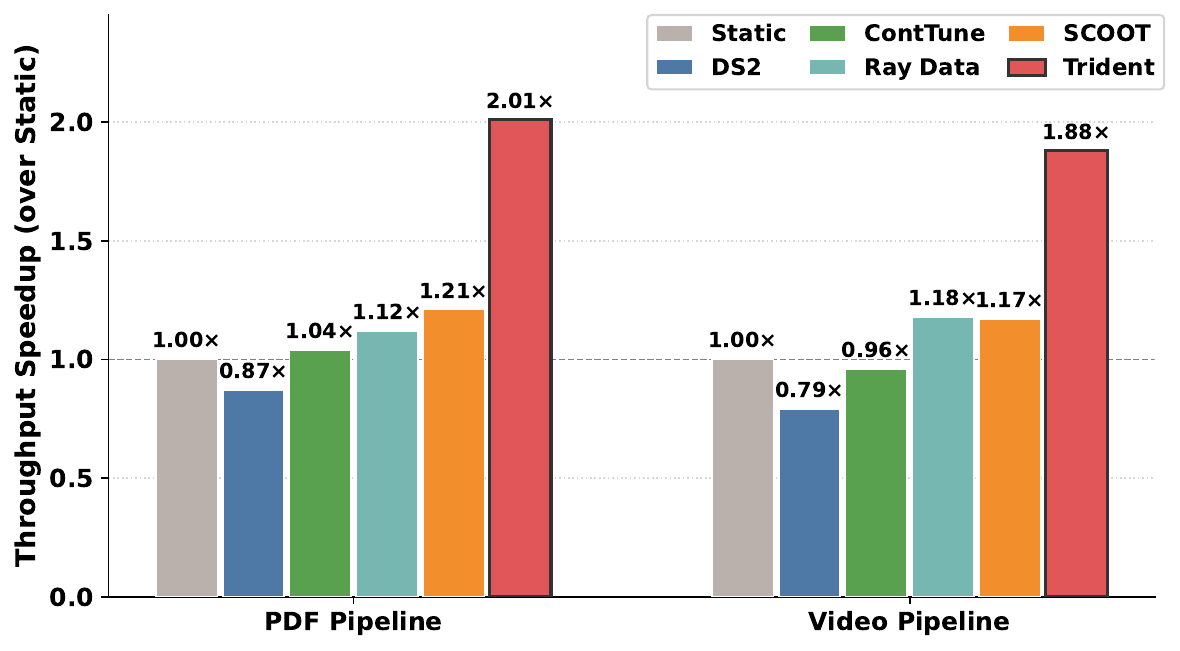}
  \caption{
  End-to-end throughput comparison on the PDF processing and video curation pipelines.
  Methods differ in their system-level coverage of capacity estimation, configuration tuning,
  and resource scheduling (Table~\ref{tab:coverage}).
  \system integrates all three layers in a closed loop.
  Speedup is reported relative to the Static baseline.
}
  \label{fig:e2e}
  %\vspace{-1.0em}
\end{figure}

We evaluate \system as a complete system against baselines under realistic deployment conditions. Each baseline covers only a subset of the three subproblems---capacity estimation, configuration tuning, and resource scheduling---while \system addresses all three through its integrated architecture, enabling coordinated decisions across components. Table~\ref{tab:coverage} summarizes each approach's coverage.

%\vspace{0.25em}
\noindent
\textbf{Setup.}
Static uses a manually tuned fixed allocation with no runtime adaptation. Ray Data~\cite{ray_data} uses its default autoscaling policy, which scales each operator independently based on in-flight tasks and utilization thresholds. DS2~\cite{ds2} estimates per-operator processing rates via useful-time instrumentation and derives parallelism from dataflow topology, assuming synchronous execution. ContTune~\cite{conttune} extends DS2 with Conservative Bayesian Optimization to learn non-linear parallelism--throughput relationships, but inherits the same useful-time instrumentation and per-operator optimization without global resource awareness. SCOOT~\cite{scoot} applies Bayesian optimization to tune inference engine parameters for each LLM-based operator offline; we run a separate tuning session per operator and deploy the resulting configurations statically with the same resource allocation as Static. \system uses the full three-layer architecture described in Sections~\ref{sec:observation}--\ref{sec:scheduling}.

%\vspace{0.25em}
\noindent
\textbf{Results.}
In Figure~\ref{fig:e2e}, \system achieves $2.01\times$ and $1.88\times$ throughput over Static on the PDF and video pipelines, respectively, outperforming all baselines.
DS2 ($0.87\times$ / $0.79\times$) and ContTune ($1.04\times$ / $0.96\times$) rely on useful-time instrumentation that  misestimates capacity for asynchronous LLM operators with continuous batching. The resulting scheduling decisions actively misallocate resources, causing both to underperform or barely match Static. Ray Data ($1.12\times$ / $1.18\times$) outperforms both by avoiding capacity modeling entirely---its threshold-based reactive scaling cannot be misled by inaccurate estimates, and thus tracks actual demand more reliably.
SCOOT ($1.21\times$ / $1.17\times$) is the strongest baseline on both pipelines, confirming that operator configuration tuning materially contributes to end-to-end throughput. However, SCOOT tunes each operator offline in isolation without accounting for inter-operator interactions, produces fixed configurations that cannot adapt to runtime workload shifts, and performs no cross-operator resource scheduling. These limitations explain \system's remaining $1.66\times$ and $1.61\times$ advantage over SCOOT: \system's observation layer provides accurate runtime capacity estimates, its adaptation layer continuously re-tunes configurations to track workload drift, and its scheduling layer jointly optimizes parallelism, placement, and configuration transitions across all operators under global resource constraints.

\begin{table}[t]
  \centering
  \caption{Subproblem coverage of evaluated approaches.}
  %\vspace{-1.0em}
  \label{tab:coverage}
  \small
  \begin{tabular}{lccc}
    \toprule
    & Observation & Adaptation & Scheduling \\
    \midrule
    Static   &              &              &              \\
    Ray Data &              &              & \checkmark   \\
    DS2      & \checkmark   &              & \checkmark   \\
    ContTune & \checkmark   &              & \checkmark   \\
    SCOOT    &              & \checkmark   &              \\
    \system  & \checkmark   & \checkmark   & \checkmark   \\
    \bottomrule
  \end{tabular}
  %\vspace{-1.0em}
\end{table}

%\vspace{-0.5em}
\subsection{Scheduling Layer Comparison (RQ2)}
\label{sec:eval:scheduling}

We isolate the scheduling layer by holding observation and adaptation constant for all methods; throughput differences are thus due only to scheduling.

%\vspace{0.25em}
\noindent
\textbf{Controlled Setup.}
All methods share (\underline{\textbf{i}}) \system's observation layer capacity estimates and
(\underline{\textbf{ii}}) \system's adaptation layer, which detects workload regime shifts and outputs
per-operator configuration recommendations.
Since baselines do not natively support configuration-transition planning, we
equip \emph{all baselines} with the same minimal mechanism to apply
recommendations: an \emph{all-at-once} switch that restarts all instances of an
affected operator simultaneously (restart overhead included in throughput).

%\vspace{0.25em}
\noindent
\textbf{Fairness Ablation.}
To separate the benefit of rolling updates from \system's global optimization,
we add \textsc{\system (all-at-once)}, which is identical to \system except it
also applies configuration changes via the same all-at-once restart.

\begin{table}[t]
  \centering
  \caption{Scheduling comparison under identical Observation+Adaptation inputs.
  Throughput normalized to Static.}
  %\vspace{-1.0em}
  \label{tab:scheduling}
  \small
  \begin{tabular}{lcc}
    \toprule
    \textbf{Method} & \textbf{PDF} & \textbf{Video} \\
    \midrule
    Static     & 1.00\(\times\) & 1.00\(\times\) \\
    Ray Data   & 1.22\(\times\) & 1.30\(\times\) \\
    DS2        & 1.38\(\times\) & 1.25\(\times\) \\
    ContTune   & 1.42\(\times\) & 1.36\(\times\) \\
    \midrule
    \system (all-at-once) & 1.92\(\times\) & 1.79\(\times\) \\
    \system               & \textbf{2.01\(\times\)} & \textbf{1.88\(\times\)} \\
    \bottomrule
  \end{tabular}
\end{table}

%\vspace{0.25em}
\noindent
\textbf{Key Findings.}
As shown in Table~\ref{tab:scheduling}, \system achieves the highest
throughput, even when all methods share identical observation and configuration
inputs.
The fairness ablation reveals that rolling updates contribute a moderate
improvement of approximately \(5\%\) over all-at-once switching on both
pipelines, confirming that controlled transitions reduce disruption during
regime changes.
However, the dominant source of \system's advantage is its globally coordinated, placement-aware scheduling.
Even under the same all-at-once switching semantics as baselines,
\textsc{\system (all-at-once)} outperforms the best baseline ContTune by
\(1.35\times\) on the PDF pipeline and \(1.32\times\) on the video pipeline,
demonstrating that joint optimization of parallelism and placement under
cluster-wide constraints---rather than a privileged transition
mechanism---accounts for most of the throughput gain.

\subsection{Observation Layer Accuracy (RQ3)}

\label{sec:eval:observation}

\begin{table}[t]
  \caption{Processing capacity estimation accuracy (MAPE~\%).}
  %\vspace{-1.0em}
  \label{tab:mape}
  \centering
  \begin{tabular}{lcc}
    \toprule
    Method & PDF & Video \\
    \midrule
    True Processing Rate & 62.7 & 54.3 \\
    EMA & 28.3 & 25.7 \\
    GP w/o filtering & 24.3 & 21.8 \\
    GP + signal filtering & 8.4 & 7.1 \\
    \textbf{GP + two-stage filtering (\system)} & \textbf{5.6} & \textbf{4.8} \\
    \bottomrule
  \end{tabular}
\end{table}

We evaluate how accurately the observation layer estimates each operator's
\emph{sustainable} throughput during end-to-end pipeline execution, where
measurements are subject to upstream starvation, downstream backpressure,
and transient imbalance.

%\vspace{0.25em}
\noindent
\textbf{Setup.}
We execute both pipelines under representative workloads and continuously
collect per-operator observations. Identical samples are fed to each
compared estimator so that accuracy differences are attributable solely
to methodology. Ground-truth capacity is obtained by profiling every
operator in isolation under sustained full-load conditions with the same
configuration as in the pipeline. Table~\ref{tab:mape} reports Mean
Absolute Percentage Error (MAPE).

%\vspace{0.25em}
\noindent
\textbf{Key Findings.}
Table~\ref{tab:mape} shows the true processing rate estimator
exhibits the highest error (62.7\%/54.3\%), confirming that synchronous
useful-time assumptions are violated by asynchronous operators with
continuous batching. EMA and unfiltered GP progressively reduce error by
smoothing variance and conditioning on workload features, respectively,
but both remain sensitive to transient pipeline effects. Signal-based
filtering sharply reduces MAPE by removing non-steady-state samples, and
\system's full two-stage filtering achieves the best accuracy
(5.6\%/4.8\%) by further rejecting GP-residual outliers. These results
confirm that accurate capacity estimation for asynchronous multimodal
operators requires both workload-aware modeling and robust anomaly
filtering.

\subsection{Adaptation Layer Effectiveness (RQ4)}

We evaluate the adaptation layer along three dimensions:
(\underline{\textbf{i}})~workload clustering accuracy,
(\underline{\textbf{ii}})~configuration optimization efficiency, and
(\underline{\textbf{iii}})~OOM protection during online exploration.

%\vspace{0.25em}
\noindent
\textbf{Clustering Accuracy.}
On the PDF pipeline (three document types processed sequentially to create
regime shifts) and the video pipeline (short-form clips vs.\ long-form
videos), we compare our online clustering against offline
K-means~\cite{kmeans} and DBSCAN~\cite{DBSCAN} that have access to the
complete dataset. As shown in Table~\ref{tab:clustering}, our online algorithm correctly discovers the expected number of clusters on both
pipelines without requiring the cluster count as input, achieving purity and ARI is only marginally below the offline baselines despite operating incrementally.

\begin{table}[t]
\centering
\caption{Workload clustering accuracy.}
%\vspace{-1.0em}
\label{tab:clustering}
\small
\begin{tabular}{lcccc}
\toprule
Method & Pipeline & Clusters & Purity & ARI \\
\midrule
K-means (offline)  & PDF   & 3 & 0.97 & 0.94 \\
DBSCAN (offline)   & PDF   & 3 & 0.95 & 0.92 \\
\system (online)   & PDF   & 3 & 0.95 & 0.89 \\
\midrule
K-means (offline)  & Video & 2 & 0.98 & 0.96 \\
DBSCAN (offline)   & Video & 2 & 0.97 & 0.95 \\
\system (online)   & Video & 2 & 0.97 & 0.93 \\
\bottomrule
\end{tabular}
\end{table}

%\vspace{0.25em}
\noindent
\textbf{Configuration Optimization.}
We evaluate memory-constrained Bayesian optimization on two representative
tunable operators: TextOCR (LLM-based, PDF pipeline) and Captioning
(Qwen2.5-VL-7B, Video pipeline). The configuration space includes inference
engine parameters such as max-num-seqs, max-num-batched-tokens,
block-size, scheduler-delay-factor,
enable-chunked-prefill, and enable-prefix-caching, forming a
large mixed-type search space. Each method is given 30 evaluations under
sustained full-load conditions. We compare against Random Search
(Sobol-based~\cite{random-search}), Grid Search, and Unconstrained BO
(standard Expected Improvement without memory constraint). All BO variants
use the Mat\'{e}rn~5/2 kernel and are initialized with 5 random evaluations.
As shown in Table~\ref{tab:config-opt}, Constrained BO achieves throughput
within 1--2\% of Unconstrained BO while substantially outperforming random
and grid search, confirming that the memory feasibility constraint introduces
only a marginal throughput trade-off. Notably, Unconstrained BO's best
configurations (marked with $\dagger$) trigger OOM during sustained pipeline
execution, negating their nominal advantage.

\begin{table}[t]
\centering
\caption{Configuration optimization comparison. Throughput normalized to
default ($1.0\times$). $\dagger$~indicates the method selected an
OOM-triggering configuration as its best.}
%\vspace{-1.0em}
\label{tab:config-opt}
\small
\begin{tabular}{lcc}
\toprule
Method & TextOCR (PDF) & Captioning (Video) \\
\midrule
Default Config      & $1.00\times$ & $1.00\times$ \\
Random Search       & $1.18\times$ & $1.14\times$ \\
Grid Search         & $1.22\times$ & $1.19\times$ \\
Unconstrained BO    & $1.38\times^\dagger$ & $1.35\times^\dagger$ \\
Constrained BO (\system) & $\mathbf{1.36\times}$ & $\mathbf{1.33\times}$ \\
\bottomrule
\end{tabular}
\end{table}

%\vspace{0.25em}
\noindent
\textbf{OOM Protection.}
To quantify the end-to-end impact, we run both pipelines, comparing
Constrained~BO ($\eta = 0.6$, $\Delta_i = 2048$\,MB) against
Unconstrained~BO with identical hyperparameters and tuning budgets.
As shown in Table~\ref{tab:oom}, Constrained BO reduces OOM events by
79--82\%, with cumulative downtime dropping from 462\,s to 102\,s (PDF) and
352\,s to 68\,s (Video). The remaining events occur during the initial random
exploration phase before the memory surrogate has sufficient observations.
Despite selecting nominally conservative configurations, Constrained BO
achieves 5.5\% (PDF) and 4.5\% (Video) higher effective throughput than
Unconstrained BO due to avoided restart overhead, with only a ${\sim}3\%$
gap relative to an OOM-free oracle.

\begin{table}[t]
\centering
\caption{OOM events and throughput impact during end-to-end pipeline execution.}
%\vspace{-1.0em}
\label{tab:oom}
\small
\begin{tabular}{lcccc}
\toprule
 & \multicolumn{2}{c}{PDF Pipeline} & \multicolumn{2}{c}{Video Pipeline} \\
\cmidrule(lr){2-3} \cmidrule(lr){4-5}
Metric & Unconstr. & Constr. & Unconstr. & Constr. \\
\midrule
OOM events              & 14    & 3     & 11    & 2     \\
Cumulative downtime (s) & 462   & 102   & 352   & 68    \\
Throughput loss vs.\ oracle & 8.7\% & 3.2\% & 7.2\% & 2.7\% \\
\bottomrule
\end{tabular}
\end{table}

\subsection{Ablation Study (RQ5)}
\label{sec:ablation}

% \begin{table}[t]
% \centering
% \caption{Ablation study results.}
% \label{tab:ablation}
% \begin{tabular}{lcc}
% \toprule
% Configuration & Execution Time & Slowdown \\
% \midrule
% \textbf{\system (Full)} & \textbf{45''29} & \textbf{---} \\
% w/o Observation Layer & 1'08''27 & +50.3\% \\
% w/o Adaptation Layer & 57''08 & +25.9\% \\
% \bottomrule
% \end{tabular}
% \end{table}

% \begin{table}[t]
% \centering
% \caption{Ablation study results.}
% \label{tab:ablation}
% \begin{tabular}{lcc}
% \toprule
% Configuration & Throughput (rows/s) & vs.\ \textsc{\system} \\
% \midrule
% \textbf{\system (Full)}   & \textbf{5.86} & \textbf{baseline} \\
% w/o Observation Layer      & 3.90          & $-$33.6\% \\
% w/o Adaptation Layer       & 4.67          & $-$20.4\% \\
% \bottomrule
% \end{tabular}
% \end{table}

% Table~\ref{tab:ablation} quantifies each component's contribution by measuring end-to-end throughput on the same workload.

% Removing the Observation Layer and falling back to traditional processing-rate estimation reduces throughput by 33.6\%. Without accurate runtime profiling, the scheduler operates on stale or inaccurate capacity estimates, leading to suboptimal parallelism decisions and persistent bottlenecks.

% Removing the Adaptation Layer and fixing the configuration throughout execution reduces throughput by 20.4\%. The degradation concentrates at regime transitions: when workload characteristics shift, the static configuration---optimized for the initial regime---misallocates resources, over-provisioning operators that are no longer saturated while starving those facing increased demand.

\begin{figure}[t]
  \centering
  \includegraphics[width=0.6\linewidth]{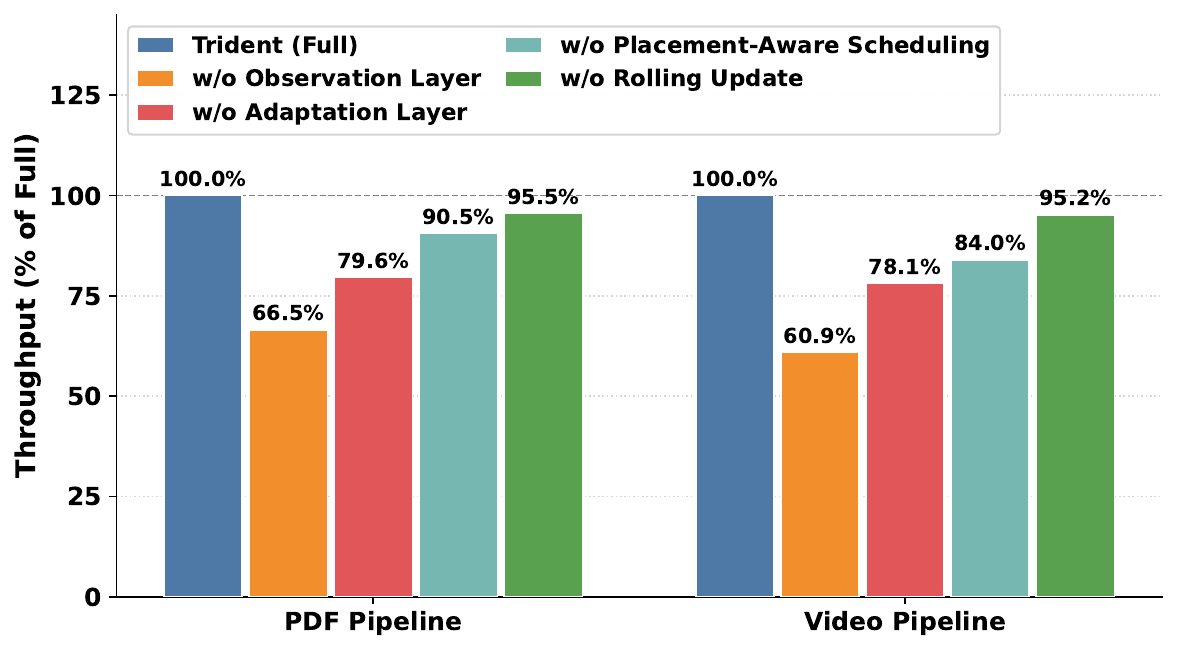}
  \caption{Ablation study. Throughput is normalized to the full \system system (100\%). Removing either layer degrades performance on both pipelines, with the observation layer contributing more in both cases.}
  \label{fig:ablation}
  %\vspace{-1.0em}
\end{figure}

We quantify the end-to-end contribution of \system's
components.

%\vspace{0.25em}
\noindent
\textbf{Setup.} We run both the \textbf{PDF} and \textbf{Video} pipelines under
the same workload traces and cluster setting as
Section~\ref{sec:exp_setup}. We report \textit{normalized throughput} where
\textit{full \system = 100\%}.

%\vspace{0.25em}
\noindent
\textbf{Variants.} We evaluate five variants: (\underline{\textbf{i}}) \textit{\system (Full)};
(\underline{\textbf{ii}}) \textit{w/o Observation Layer}, which replaces \system's GP-based
sustainable-throughput estimation with the true-processing-rate estimator
based on useful time; (\underline{\textbf{iii}}) \textit{w/o Adaptation Layer}, which disables online
workload clustering and configuration tuning so operators use fixed initial configurations; (\underline{\textbf{iv}}) \textit{w/o Placement-Aware Scheduling}, which makes the
scheduling optimization \emph{network-agnostic} by removing network/
communication modeling and co-location incentives while keeping global
parallelism and configuration-transition optimization unchanged; and
(\underline{\textbf{v}}) \textit{w/o Rolling Update}, which applies each configuration
change by restarting all instances of the affected operator simultaneously.

%\vspace{0.25em}
\noindent
\textbf{Results.} Figure~\ref{fig:ablation} shows that the observation layer is the most critical component: removing the observation layer reduces throughput to
\textbf{66.5\%} (PDF) and \textbf{60.9\%} (Video), indicating that accurate
sustainable-capacity estimation is essential for correct end-to-end resource
allocation in pipelines with asynchronous and dynamically batched operators.
Disabling the adaptation layer decreases throughput to \textbf{79.6\%}
(PDF) and \textbf{78.1\%} (Video), showing that online workload-aware
configuration tuning provides substantial gains beyond scheduling alone.
Removing \textit{placement awareness} leads to a smaller but consistent
drop---to \textbf{90.5\%} (PDF) and \textbf{84.0\%} (Video)---highlighting the
importance of co-locating communication-intensive stages, particularly for the
Video pipeline. Finally, replacing rolling updates with \textbf{all-at-once}
restarts reduces throughput to \textbf{95.5\%} (PDF) and \textbf{95.2\%}
(Video), confirming that controlled transitions mitigate cold-start disruption
during configuration switches.

\subsection{System Overhead (RQ6)}

The observation layer and adaptation layer execute synchronously within Ray Data's scheduler loop, adding only 2\,ms and 4\,ms per invocation respectively---negligible compared to the scheduler's own approximately ${\sim}400$\,ms execution time. The scheduling layer's MILP is solved asynchronously by a dedicated Ray actor, so the pipeline continues under the most recent feasible solution while a new one is computed. On our production cluster of 8 nodes, the mean MILP solve time is 206\,ms (PDF pipeline) and 62\,ms (video pipeline). When scaling the cluster to 16 nodes, solve time grows to 1{,}521\,ms and 259\,ms respectively, still well within the multi-minute rescheduling interval and entirely off the critical path.

\section{Related Work}

\textbf{Dataflow Scheduling and Stream Processing Autoscaling.}
Dataflow systems such as Flink~\cite{flink}, Spark Streaming~\cite{spark-streaming}, and Ray Data~\cite{ray_data} regulate execution via backpressure or utilization thresholds. Autoscaling strategies range from rule-based policies~\cite{distributed_operation,scale_out_and_fault_tolerance,dhalion,elastic_scaling,stream_cloud,latency-aware,stela,spark-streaming}, to model-based methods like DS2~\cite{ds2} and Turbine~\cite{Turbine} that derive parallelism from useful-time measurements, to learning-based techniques such as Dragster~\cite{Dragster} and ContTune~\cite{conttune} that apply Bayesian optimization. These approaches uniformly assume elastic resources, synchronous operators, and negligible transfer costs. Notably, useful-time instrumentation systematically misestimates the capacity of asynchronous operators under continuous batching (Section~\ref{sec:eval:observation}). \system addresses these gaps by jointly optimizing operator parallelism, device placement, and configuration transitions under a fixed heterogeneous resource budget via a unified MILP formulation.

%\vspace{0.25em}
\noindent
\textbf{Performance Modeling with Gaussian Processes.}
GP regression is widely adopted for black-box performance modeling: OtterTune~\cite{ottertune} learns DBMS performance surfaces for knob tuning, CherryPick~\cite{cherrypick} applies GP-based BO to cloud VM selection, and Fu et al.~\cite{fu2021} show that inherent performance variability dominates prediction error, motivating robust observation filtering. Online extensions including SkyGP~\cite{skygp} and DAO-GP~\cite{daogp} handle non-stationarity through local expert maintenance and drift detection, though neither targets data pipelines with asynchronous operators. \system's observation layer builds on this line by coupling GP regression with a two-stage anomaly filter that removes starvation and backpressure artifacts, yielding sustainable capacity estimates that update incrementally as workloads evolve.

%\vspace{0.25em}
\noindent
\textbf{Constrained Configuration Tuning and Safe Exploration.}
Constrained Bayesian optimization has been applied extensively to database tuning: ResTune~\cite{restune} models both objectives and SLA constraints with separate GPs, OnlineTune~\cite{onlinetune} restricts exploration to a GP-derived safety region for live databases, LlamaTune~\cite{llamatune} reduces dimensionality via randomized embeddings, and QTune~\cite{qtune} introduces query-aware encodings for robustness under workload drift. In the LLM inference domain, SCOOT~\cite{scoot} tunes serving parameters per operator using BO with learned hidden constraints. \system's adaptation layer shares the constrained BO formulation but targets peak accelerator memory---whose violation triggers OOM failures and prolonged restarts---and couples configuration search with online workload regime discovery, forwarding per-regime recommendations to the scheduling layer's MILP solver for transition-aware commitment decisions.

\section{Conclusion}
In this paper, we presented \system, an adaptive scheduling framework for heterogeneous multimodal data preparation pipelines running on fixed-resource clusters. \system addresses key limitations of existing stream processing schedulers when applied to modern multimodal workloads, including inaccurate throughput estimation for asynchronous operators, static operator configurations under workload heterogeneity, and the lack of placement-aware resource allocation.
\system integrates three tightly coupled components: (\underline{\textbf{i}}) a noise-resilient observation layer to estimate sustainable operator throughput, (\underline{\textbf{ii}}) an adaptation layer that performs online workload clustering and memory-constrained configuration tuning, and (\underline{\textbf{iii}}) a scheduling layer that jointly optimizes operator parallelism, placement, and configuration transitions using a mixed-integer linear program, where rolling updates are selectively applied only when the projected throughput gain outweighs the cold-start overhead. We implemented \system on top of Ray Data and evaluated \system on production-representative document and video processing pipelines. Experimental results show that \system improves end-to-end throughput by up to \(2.01\times\)  compared to state-of-the-art approaches, while maintaining low overhead and stable behavior under workload regime shifts.

% \begin{acks}
%  This work was supported by the [...] Research Fund of [...] (Number [...]). Additional funding was provided by [...] and [...]. We also thank [...] for contributing [...].
% \end{acks}

\clearpage

\bibliographystyle{unsrt}
\bibliography{references}

\end{document}